\documentclass[twocolumn,twocolappendix]{aa}

\usepackage[utf8]{inputenc}
\usepackage{amsmath,amssymb,dsfont,stmaryrd,bm}
\usepackage{graphicx}
\usepackage{tabularx}
\usepackage{multirow}
\usepackage{soul,xcolor}
\usepackage{acronym}
\usepackage{hyperref}
\usepackage{times}
\usepackage{beramono}

\newcommand{\ra}{\mathrm{a}}
\newcommand{\rd}{\mathrm{d}}
\newcommand{\rD}{\mathrm{D}}
\newcommand{\re}{\mathrm{e}}
\newcommand{\ri}{\mathrm{i}}
\newcommand{\rp}{\mathrm{p}}
\newcommand{\RePart}{\mathrm{Re}}
\newcommand{\ImPart}{\mathrm{Im}}
\newcommand{\res}{\mathrm{res}}
\newcommand{\rapo}{r_{\ra}}
\newcommand{\rper}{r_{\rp}}
\newcommand{\Mtot}{M_{\mathrm{tot}}}
\newcommand{\Tin}{T_\mathrm{in}}
\newcommand{\Tout}{T_\mathrm{out}}
\newcommand{\Rin}{R_\mathrm{in}}
\newcommand{\Rout}{R_\mathrm{out}}
\newcommand{\Rmax}{R_\mathrm{max}}
\newcommand{\psieff}{\psi_{\mathrm{eff}}}
\newcommand{\tdyn}{t_{\mathrm{dyn}}}
\newcommand{\bJ}{\mathbf{J}}
\newcommand{\bk}{\mathbf{k}}
\newcommand{\bOmega}{\mathbf{\Omega}}
\newcommand{\br}{\mathbf{r}}
\newcommand{\mybtheta}{\boldsymbol{\theta}}
\newcommand{\bv}{\mathbf{v}}
\newcommand{\bw}{\mathbf{w}}

\newcommand{\bx}{\mathbf{x}}
\newcommand{\scM}{\textsc{m}}
\newcommand{\omegaM}{\omega_{\scM}}
\newcommand{\OmegaM}{\Omega_{\scM}}
\newcommand{\gammaM}{\gamma_{\scM}}
\newcommand{\mO}{\mathcal{O}}
\newcommand{\boldmatrix}[1]{\boldsymbol{\mathsf{#1}}}
\newcommand{\bI}{\boldmatrix{I}}
\newcommand{\bM}{\boldmatrix{M}}
\newcommand{\bN}{\boldmatrix{N}}
\newcommand{\p}{\partial}
\newcommand{\half}{\tfrac{1}{2}}
\newcommand{\dirac}{\delta_{\rD}}
\newcommand{\flux}{\boldsymbol{\mathcal{F}}}
\newcommand{\orbitalelements}{{\tiny\texttt{OrbitalElements.jl}}}
\newcommand{\astrobasis}{{\tiny\texttt{AstroBasis.jl}}}

\newcommand{\linearresponse}{{\tiny\texttt{LinearResponse.jl}}}
\newcommand{\julia}{{\tiny\texttt{julia}\,}}
\newcommand{\eq}{equation}
\newcommand{\eqs}{equations}
\newcommand{\fig}{figure}
\newcommand{\figs}{figures}
\newcommand{\tab}{table}

\newcommand{\sect}{section}

\newcommand{\app}{appendix}

\newcommand{\exteq}{equation}
\newcommand{\exteqs}{equations}
\newcommand{\extfig}{figure}

\newcommand{\exttab}{table}
\newcommand{\exttabs}{tables}
\newcommand{\extsect}{section}

\newcommand{\extapp}{appendix}
\newcommand{\extapps}{appendices}
\newcommand{\Nbody}{$N$--body}

\newcommand{\eps}{\epsilon}
\newcommand{\veps}{\varepsilon}
\defcitealias{Sellwood2012}{S12}
\defcitealias{Fouvry+2015}{F\texttt{+}15}
\defcitealias{DeRijcke+2019b}{DR\texttt{+}19b}
\defcitealias{Petersen+2024}{PR\texttt{+}24}
\newacro{HMF}{Hamiltonian mean field}
\newcommand{\HMF}{\ac{HMF}}
\newacro{ILR}{inner Lindblad resonance}
\newcommand{\ILR}{\ac{ILR}}
\newacro{DF}{distribution function}
\newacroplural{DF}[DFs]{distribution functions}
\newcommand{\DF}{\ac{DF}}
\newcommand{\DFs}{\acp{DF}}
\newacro{BL}{Balescu--Lenard}
\newcommand{\BL}{\ac{BL}}

\hypersetup{colorlinks=true, linkcolor=black, citecolor=black, filecolor=blue, urlcolor=black}

\newcolumntype{C}[1]{>{\centering\arraybackslash}m{#1}}

\definecolor{bubbles}{rgb}{0.91, 1.0, 1.0}
\definecolor{aquamarine}{rgb}{0.5, 1.0, 0.83}
\definecolor{bubblegum}{rgb}{0.99, 0.76, 0.8}
\definecolor{bluebell}{rgb}{0.74, 0.74, 0.92}
\definecolor{dollarbill}{rgb}{0.72, 0.93, 0.6}

\newcounter{Xtophecounter}
\newcounter{JBcounter}
\newcounter{MRcounter}

\begin{document}

\title{On the long-term evolution of razor-thin galactic discs:\\
Balescu--Lenard prediction and perspectives }
\titlerunning{On the long-term evolution of razor-thin galactic discs}

\authorrunning{M. Roule, J.~B. Fouvry, C. Pichon \& P. H. Chavanis}

\author{
Mathieu~Roule\inst{1} \and 
Jean-Baptiste Fouvry\inst{1} \and
Christophe~Pichon\inst{1,2,3}\thanks{Corresponding author: pichon@iap.fr}
\and
Pierre-Henri Chavanis\inst{4} 
}
\institute{
Institut d'Astrophysique de Paris, CNRS and Sorbonne Universit\'e, UMR 7095, 98 bis Boulevard Arago, F-75014 Paris, France
\and
IPhT, DRF-INP, UMR 3680, CEA, L'Orme des Merisiers, B\^at 774, 91191 Gif-sur-Yvette, France
\and
 Kyung Hee University, Dept. of Astronomy \& Space Science, Yongin-shi, Gyeonggi-do 17104, Republic of Korea
\and
Laboratoire de Physique Th\'eorique, Universit\'e de Toulouse, CNRS, UPS, France
}

\abstract{
In the last five decades, numerical simulations have provided invaluable insights into the evolution of galactic discs over cosmic times. As a complementary approach, developments in kinetic theory now also offer a theoretical framework to understand statistically their long-term evolution. 
The current state-of-the-art kinetic theory of isolated stellar systems is the inhomogeneous Balescu--Lenard equation. It can describe the long-term evolution of a self-gravitating razor-thin disc under the effect of resonant interactions between collectively amplified noise-driven fluctuations.
In this work, confronting theoretical predictions to numerical simulations, we quantitatively show that
kinetic theory indeed captures the average long-term evolution of cold stellar discs.
Leveraging the versatility of kinetic methods, we then offer some new perspectives on this problem,
namely (i) the crucial impact of collective effects in accelerating the relaxation;
(ii) the role of (weakly) damped modes in shaping the disc's orbital heating;
(iii) the  bias introduced by gravitational softening on long timescales;
(iv) the resurgence of strong stochasticity near marginal stability.
These elements call for an appropriate choice of 
softening kernel when simulating the long-term evolution of razor thin discs
and  for an extension of kinetic theory beyond the average evolution.
Notwithstanding, kinetic theory captures quantitatively
the ensemble-averaged long-term response of such discs.
}


\maketitle

\section{Introduction}
\label{sec:intro}

Kinetic theory aims at describing statistically the evolution of many-body systems over long timescales.
This endeavour is timely given the wealth of observational data recently available,
e.g.\@, GAIA~\citep{Gaia+2016} that offers an unprecedented view
on the dynamical state of the Milky Way~\citep[see, e.g.\@,][]{Trick+2019,Hunt+2019},
as well as large surveys such as JWST~\citep{JWST2006}
that give access to samples of  galaxies with various morphological types
across cosmic times \citep{Kuhn+2024}.
To leverage this statistical sample of galactic observations,
the goal of kinetic theory is to capture their long-term evolution via a 
 ``collision operator''
that describes quantitatively these system's  mean relaxation,
i.e., the dynamical rearrangement of their orbital  distribution.
We refer to~\cite{Chavanis2013stellar,Chavanis2024}
for a thorough historical account on kinetic theory applied to self-gravitating systems.
In this work, we will focus on the use of kinetic theory
to describe the self-induced relaxation of galactic discs,
a key ingredient of their morphological transformation.

If one's goal is to describe the long-term evolution of a stellar disc,
a few key properties must be accounted for.
First, galactic discs are spatially \textit{inhomogeneous},
i.e., stars follow intricate orbits,
as described by angle-action coordinates~\citep{BinneyTremaine2008}.
Second, galactic discs are \textit{resonant},
i.e., stellar orbits may resonate with one another,
e.g.\@, to source dynamical friction~\citep{LyndenBellKalnajs1972}.
Third, galactic discs are \textit{self-gravitating},
i.e., discs strongly enhance perturbations,
e.g.\@, through swing amplification arising from collective effects~\citep{JulianToomre1966}.
Fourth, galactic discs are \textit{discrete},
i.e.\@, they are composed of a finite number of constituents,
hence unavoidably submitted to an intrinsic Poisson fluctuations.

It is only recently that a fully self-consistent kinetic theory managed
to take into account all these defining features.
It is the celebrated \textit{inhomogeneous Balescu--Lenard equation}~\citep{Heyvaerts2010,Chavanis2012}.
This kinetic equation fully embraces spatial inhomogeneity
to describe the average long-term impact of resonantly-coupled, collectively amplified and internally-driven
fluctuations on the orbital distribution of a stellar system.
Phrased differently, this master equation, on paper at least,
encompasses all the key dynamical properties of isolated stellar discs.
Since its derivation, this equation has been successfully and quantitatively
applied to various systems such as the \HMF\@ model~\citep{BenettiMarcos2017}, galactic nuclei~\citep{FouvryBarOr2018}, globular clusters~\citep{Fouvry+2021} and one-dimensional gravity~\citep{Roule+2022}.
Yet its validation on cold galactic discs has remained qualitative  at best~\citep{Fouvry+2015}.

This paper addresses two key questions regarding the fate of isolated stellar discs:
(i) How do resonant interactions and collective effects shape the long-term evolution of stellar discs?
(ii) What are the limitations of kinetic theories in predicting the evolution of these cold self-gravitating systems?
For that purpose, we study the Mestel disc~\citep{ZangThesis}
as the testbed of kinetic theory.

This paper is organised as follows.
Section~\ref{sec:model_review} presents the disc model we consider and reviews previous works.
Section~\ref{sec:long-term} then considers the disc's averaged long-term evolution
as measured in \Nbody\ simulations and predicted by the \BL\@ equation.
Section~\ref{sec:discussion} 
discusses in turn the impact of self-gravity,
damped modes,
softening
and long-term stochasticity near phase transition.
Finally, \sect~\ref{sec:conclusion} wraps up and offers some perspectives
for future investigations.
Throughout the main text, technical details are kept to a
minimum and deferred to Appendices or to relevant references.

\section{Model and earlier results}
\label{sec:model_review}

\subsection{Razor-thin disc model}
\label{sec:model}

We focus on tapered Mestel discs \citep{10.1093/mnras/126.6.553},
whose dynamics has been extensively studied~\citep[e.g.,][]{ZangThesis,Toomre1981,EvansRead1998,SellwoodEvans2001,Sellwood2012,Fouvry+2015}.
A razor-thin Mestel disc has constant circular velocity, 
$V_{0}$,
to mimic the (relatively) flat rotation curve of the Milky Way~\citep[see, e.g.\@,][]{Eilers+2019}.
The associated potential is
\begin{equation}
\label{eq:Mestel_potential}
\psi(r) = V_0^2 \ln (r/R_0),
\end{equation}
which sets the dynamical time to ${ \tdyn \!=\! V_0/ R_0 }$.
In the following, we work within the units ${G\!=\!R_0\!=\!V_0\!=\!1}$.
Mestel discs have infinite mass and their orbital frequencies diverge like ${1/r}$ in the centre.\footnote{\label{fn:soft_pot}To ease the numerics, we soften \eq~\eqref{eq:Mestel_potential}
with ${ r / R_{0} \!\to\! \sqrt{(r/R_{0})^{2} \!+\! \eps^{2}} }$,
with ${\eps \!\ll\! 1}$. We checked that this softening had no impact on our predictions.}
A compatible \DF\@ for a Mestel disc is given by~\citep[see, e.g.\@,][]{Chakrabarty+2024}
\begin{equation}
\label{eq:Mestel_DF}
F(E,L) = C \, L^q \, \re^{-E/\sigma^2},
\end{equation}
where $\sigma$, the radial velocity dispersion,
controls the disc's dynamical temperature,
${ q \!=\! q (\sigma) }$,
and $C$ is a normalisation constant.
Finally, to mimic the disc's inner bulge
and make the disc's total mass finite,
we also introduce an inner and an outer tapering in the distribution,
as detailed in \app~\ref{app:disc_DF}.

The present study is motivated by the work of~\citet[][hereafter~\citetalias{Sellwood2012}]{Sellwood2012}
and~\citet[][hereafter~\citetalias{Fouvry+2015}]{Fouvry+2015}
which investigated a particular razor-thin model of this family (see \app~\ref{app:disc_DF} for detailed parameters).
\citetalias{Sellwood2012} used \Nbody\@ simulations,
while \citetalias{Fouvry+2015} implemented \BL\@.
Let us now revisit their main result
before delving further.

\subsection{\protect\cite{Sellwood2012}'s \Nbody\@ results}
\label{sec:longterm_discs_sellwood12}

\citetalias{Sellwood2012} performed long-term simulations
of a razor-thin Mestel disc,
limiting themselves solely to the ${ \ell \!=\! 2 }$ fluctuations.
In their \extfig~{2}, \citetalias{Sellwood2012} showed that,
albeit initially stable, the disc becomes linearly unstable,
and that the time at which the instability sets in depends on the number of particles, $N$.
To prove that the disc was indeed unstable at late times,
\citetalias{Sellwood2012} stopped the simulations at different times,
and restarted them after reshuffling the stars' orbital phases.
Even if this killed, de facto, any coherent bisymmetric feature,
the disc was still developing a clear instability (\extfig~{5} therein).
The later the reshuffling, the stronger the instability.

Such a transition from a stable to an unstable distribution
cannot be explained by linear theory. It necessarily involves changing
the disc's mean \DF\@.
When looking at the distribution of orbits in action space, 
\citetalias{Sellwood2012} reported on
(i) a localised depletion of circular orbits (or groove) before the instability kicks in (\extfig~{10} therein);
(ii) the presence of a strong, sharp ridge at resonance with the instability before it saturates (\extfig~{8} therein).
The groove generated during the step (i) is responsible for the nascent instability in step (ii)~\citep{SellwoodKahn1991,DeRijcke+2019a}.

More recently, \cite{Sellwood2020} confirmed that their result
were left unchanged when changing the \Nbody\ code from a polar to a Cartesian grid.
Different individual realisations were also giving similar evolution tracks for the fluctuations.\footnote{In both cases, the initial conditions were from a ``quiet start''~\citep{DebattistaSellwood2000,Sellwood2024}.
It would be interesting to investigate
whether such non-Poisson-like initial conditions
can bias the disc's long-term relaxation.}
The slow growth of the fluctuations during the onset of relaxation
was still to be explained.

\subsection{\protect\cite{Fouvry+2015}'s results}
\label{sec:longterm_discs_fouvry15}

\citetalias{Fouvry+2015} investigated \citetalias{Sellwood2012}'s disc
by running their own numerical simulations
as well as by implementing  the corresponding kinetic theory.
With their simulations, \citetalias{Fouvry+2015} showed that (i) the disc's early relaxation happens on a timescale proportional to $N$;
(ii) increasing the disc's active fraction, $\xi$ (\eq~\ref{eq:tapered_Mestel_DF}), accelerates the relaxation, in a non-linear fashion.
Then, implementing the inhomogeneous \BL\@ \eq~\eqref{eq:BL},
\citetalias{Fouvry+2015} found that relaxation of the \DF\@ predicted by \BL\
was qualitatively consistent with the measurement in one \Nbody\@ simulation,
exhibiting a ridge in action space.
Importantly, they emphasised how collective amplification
considerably reshapes the properties of the relaxation.

Taken together, \citetalias{Sellwood2012} and \citetalias{Fouvry+2015} offer a convincing picture for the long-term evolution of razor-thin discs.
Initially stable, the disc first evolves under the collisional effects of swing amplified finite-$N$ fluctuations.
This long-term relaxation is characterised by a strong heating and churning \citep{SellwoodBinney2002} of circular orbits.
Ultimately, the new \DF\@ becomes linearly unstable
allowing for an exponential growth of fluctuations.

Yet, a few potential caveats in this story can be raised and deserve further investigation.
\begin{enumerate}
\item The ridge presented in \citetalias{Sellwood2012}'s simulation is measured when the instability is already present.
This ridge might not correspond to the initial change in the \DF\@
but rather be induced by a new instability.
Measurements at earlier time~\citepalias[see][\extfig~{10} therein]{Sellwood2012}
show a much broader depletion of circular orbits.
\item \BL\@ (\eq~\ref{eq:BL}) predicts the mean evolution of the \DF\@ averaged over different realisations.
\BL\ does not predict the evolution of a single (quiet start) \Nbody\@ realisation,
i.e., what was measured by~\citetalias{Sellwood2012}.
\item The agreement between \citetalias{Sellwood2012}'s \Nbody\ measurements and \citetalias{Fouvry+2015}'s prediction is more qualitative than quantitative. The predicted ridge and the measured one are not at the same location
nor with a precisely matching amplitude.
\end{enumerate}
Let us now rely on recent improvements to numerical kinetic theory and our own set of \Nbody\@ simulations to elucidate these elements.

\section{Long-term evolution}
\label{sec:long-term}

\subsection{Balescu--Lenard equation}
\label{sec:BL}

The long-term relaxation of self-gravitating stellar systems driven by finite-$N$ fluctuations
is governed by the inhomogeneous \BL\ equation~\citep{Heyvaerts2010,Chavanis2012}. It reads
\begin{align}
\frac{\p F (\bJ, t)}{\p t}& \!=\! - \pi (2\pi)^{d} \, m \, \frac{\p}{\p\bJ} \!\cdot\!
\!\bigg[\sum_{\bk, \bk'} \bk \!\! \int \!\! \rd \bJ' 
\underbrace{\big| U^{\rd}_{\bk\bk'} (\bJ, \bJ', \bk\!\cdot\!\bOmega) \big|^{2}}_{\text{dressed coupling}}\notag \\
& \hskip -1.4cm\times {} 
\underbrace{\dirac ( \bk\!\cdot\!\bOmega \!-\! \bk'\!\cdot\!\bOmega' ) }_{\text{resonance condition}}
\underbrace{\left(\!\bk'\!\cdot\!\frac{\p}{\p \bJ'}\right.}_{\text{friction}}
\!-\!\underbrace{\left.\bk\!\cdot\!\frac{\p}{\p\bJ} \!\right)}_{\text{diffusion}}
\underbrace{F (\bJ, t) \, F(\bJ', t)}_{\text{orbital population}}\bigg] ,
\label{eq:BL}
\end{align}
with ${ \bOmega \!=\! \bOmega (\bJ) }$, the vector of orbital frequencies, and similarly ${ \bOmega' \!=\! \bOmega (\bJ') }$.
This is the master equation of self-induced orbital relaxation
This non-linear equation describes how the mean orbital population distribution ${F(\bJ, t)}$,
with $\bJ$ the action (\app~\ref{app:disc_AA}), evolves through the correlated effects of Poisson noise, i.e., finite-${N}$ effects (with ${m\!\propto\!1/N}$ the stars' individual mass).
Importantly, \eq~\eqref{eq:BL} conserves mass, energy, and satisfies an $H$-theorem for Boltzmann entropy \citep{Heyvaerts2010}.
It captures the small but cumulative effects of resonant encounters between stars,
whose efficiency is dressed by collective effects.
The sum and the integral in this equation scan over the discrete resonances, ${ (\bk , \bk') }$,
and over the orbital space, ${ \!\int\! \rd \bJ' }$, looking for all the possible populated resonances. 
These resonances are selected through the resonance condition, ${\bk\!\cdot\!\bOmega\!-\!\bk'\!\cdot\!\bOmega'\!=\!0}$.

In \eq~\eqref{eq:BL},
the efficiency of a given resonant coupling is set by the dressed coupling coefficients,
${ U^{\rd}_{\bk\bk'} }$.
These coefficients encompass all the details of linear response theory.
This is briefly reviewed and validated in \app~\ref{app:linear}.
As the upcoming sections show, these coefficients play an extremely important role
in defining the disc's relaxation.
Once linear response is under control,
one may proceed with the evaluation of the \BL\ relaxation rate,
as detailed in \app~\ref{app:BL}.
Naturally, the dynamics of the Mestel disc can also be investigated
using \Nbody\ simulations. We detail our setup in \app~\ref{app:Nbody}.

\subsection{Dynamical phase transition}
\label{sec:DynamicalPhaseTransition}

\begin{figure}
\begin{center}
\includegraphics[width=0.48\textwidth]{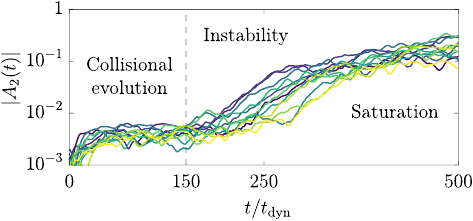}
\caption{Time evolution of the bisymmetric fluctuations (\eq~\ref{eq:mode_projection}) in 12 different \Nbody\@ realisations of the Mestel disc, with ${ N \!=\! 25 \!\times\! 10^{6} }$ particles each,
using a running average over $30$ dynamical times.
The disc is initially stable and slowly relaxes towards an unstable state.
Once unstable, the evolution is dominated by an exponentially growing mode before it saturates.
The disc's configuration is illustrated in \fig~\ref{fig:disc_configuration}.
Interestingly, the dispersion among realisations greatly increases close to the instability.
The dashed line at ${t/\tdyn\!=\!150}$ is the time at which the changes in the \DF\@ are measured in \fig~\ref{fig:discs_BL_NBODY}.
}
\label{fig:discs_fluctuations}
\end{center}
\end{figure}

\begin{figure}
    \begin{center}
    \includegraphics[width=0.48\textwidth]{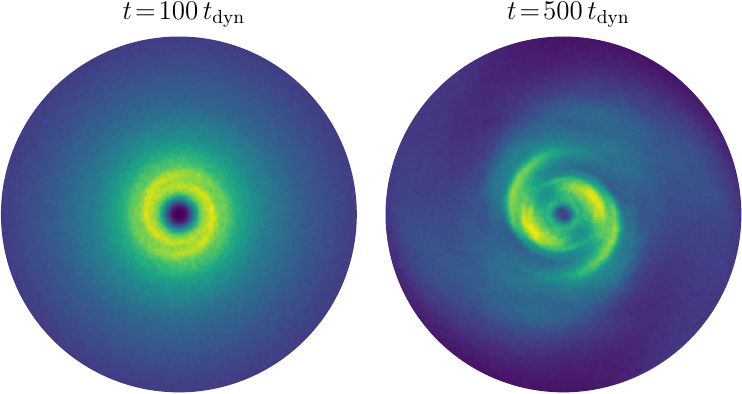}
    \caption{Illustration of the typical evolution of a Mestel disc in one \Nbody\ simulation,
    via the surface density (up to ${R\!=\!8R_0}$) at two different times.
    During the collisional phase (left), the disc remains reasonably axisymmetric,
    displaying recurrent weak transient spirals.
    After the phase transition (right), the disc develops strong bisymmetric fluctuations,
    which ultimately saturate (\fig~\ref{fig:discs_fluctuations}),
    up to the late formation of a bar~\citepalias[see \extfig~{10} in][]{Fouvry+2015}.
    }
    \label{fig:disc_configuration}
\end{center}
\end{figure}

As a first inspection of this disc's evolution,
we present in \fig~\ref{fig:discs_fluctuations},
the time evolution of bisymmetric fluctuations
in the disc, as observed in \Nbody\ simulations.
In that figure, different lines correspond to
different realisations of the same disc,
i.e., they only differ in their initial conditions.
First, we can note that for ${ t \!\lesssim\! 200 \, \tdyn }$,
the evolution of the discs seems rather smooth and quiescent.
This is a phase of ``collisional evolution'',
driven by finite-$N$ fluctuations and described by \BL\@.
During this period, the \DF\ changes with time
but remains dynamically (Vlasov) stable.
Then, for ${ t \!\gtrsim\! 200 \, \tdyn }$,
the \DF\ becomes dynamically (Vlasov) unstable because of the groove formation,
the discs change of evolution regime,
and the growth of the fluctuations becomes much more rapid.
The discs have gone through a dynamical phase transition toward instability~\citep[see, e.g.\@,][for similar behaviour in another long-range interacting system]{Campa+2008}.
Finally, we note that the dispersion among the different realisations gets much larger as the phase transition is approached.
In \fig~\ref{fig:disc_configuration}, we illustrate
one disc simulation in these two regimes.

\subsection{Ensemble-averaged relaxation rate}
\label{sec:EnsembleAveragedRelaxationRate}

\begin{figure}
\begin{center}
\includegraphics[width=0.48\textwidth]{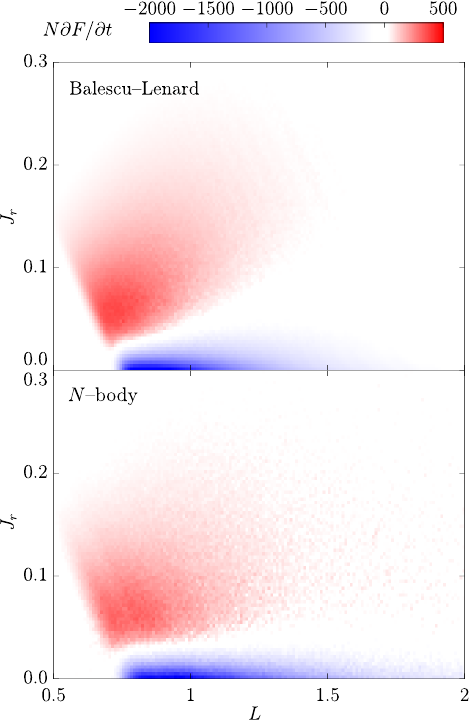}
\caption{\textit{Top}: Local relaxation rate, $\p F/\p t$, in action space as predicted by \BL\@ (\eq~\ref{eq:BL}) for the Mestel disc (\sect~\ref{sec:model}),
computed in the centre of the action bins.
Red regions correspond to an increase in the number of particles while blue contours correspond to a depletion.
\textit{Bottom}: Relaxation rate measured in \Nbody\@ simulations of the same disc,
averaged over ${1\,000}$ realisations with ${N\!=\!25\!\times\!10^6}$ particles each.
The changes in the \DF\@ are computed at ${t/\tdyn\!=\!150}$,
safely before the discs become unstable (\fig~\ref{fig:discs_fluctuations}).
The prediction and measurement are in good agreement, both in shape and amplitude.
Slices in action space are illustrated in \fig~\ref{fig:slices}.
Collective effects play a crucial role in shaping the long-term evolution of razor-thin discs. 
}
\label{fig:discs_BL_NBODY}
\end{center}
\end{figure}

Let us now focus on the first stage of evolution and consider
the averaged relaxation rate in action space, ${ \p F (\bJ , t) / \p t }$,
at the time ${ t \!=\! 150 \, \tdyn }$.
This is what we present in \fig~\ref{fig:discs_BL_NBODY},
one of the main result of the present work.
In that figure,
we compare the \BL\ prediction of the relaxation rate (top panel)
with the ensemble-averaged relaxation rate measured in \Nbody\ simulations.
As expected, we recover that discs are heated up by resonant encounters,
and orbits diffuse from quasi-circular orbits to more eccentric ones.
We find that the kinetic prediction and the \Nbody\ measurements are in quantitative agreement.
Both approaches predict similar shape for the relaxation rate in action space,
and, importantly, with similar amplitude.
Using the same criterion as in~\exteq~{(12)} of~\cite{Tep+2022},
we find that ${  \!\int\! \rd \bJ \, F \, |\p F / \p t | }$ agrees within $10\%$ between both panels.
This is an important validation of \BL\@.

\begin{figure}
\begin{center}
\includegraphics[width=0.48\textwidth]{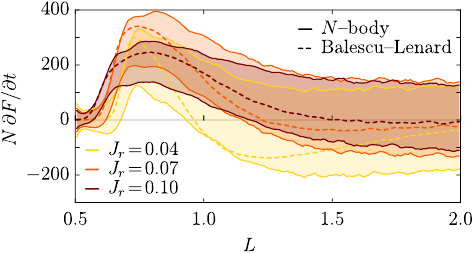}
\includegraphics[width=0.48\textwidth]{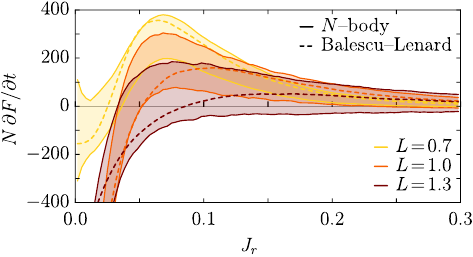}
\caption{Slices of the local relaxation rate, ${ \p F / \p t }$, from \fig~\ref{fig:discs_BL_NBODY} for fixed $J_{r}$ (top) or fixed $L$ (bottom),
as predicted from \BL\ (dashed) and measured in \Nbody\ simulations (full lines).
Here, ${ \p F / \p t }$ in the top (resp.\ bottom) panel has been averaged over an interval of width ${ \delta J_{r} \!=\! \pm 0.02 }$ (resp.\ ${ \delta L \!=\! 0.06 }$) and subsequently smoothed with a running average
of width ${ \delta L \!=\! 0.03 }$ (resp.\ ${ \delta J_{r} \!=\! 0.01 }$).
For the \Nbody, we measured ${ \p F / \p t }$ in each realisation independently,
estimated the standard deviation among the sample of realisations,
and represented the level lines one standard deviation away from the mean value.
}
\label{fig:slices}
\end{center}
\end{figure}

To conclude, \fig~\ref{fig:slices}
 presents slices of \fig~\ref{fig:discs_BL_NBODY} through action space.
These slices show that the qualitative agreement seen in \fig~\ref{fig:discs_BL_NBODY} is indeed quantitative.
Indeed, the amplitude of the \BL\ predictions 
falls within the distribution of simulated relaxation rates.
This figure also highlights the large spread
in the relaxation rate already visible in \fig~\ref{fig:discs_fluctuations}.
Such an agreement between \BL\@ and \Nbody\@ simulations was definitely not a given.
Indeed, the present discs are close to being marginally stable.
This corresponds to a dynamical regime where strong collective amplification
via the contributions from damped modes
may have a significant impact on long-term relaxation~\citep[see, e.g.\@,][]{HamiltonHeinemann2020}.

\section{Discussion}
\label{sec:discussion}

Following \fig~\ref{fig:discs_BL_NBODY},
one could be tempted to conclude that \BL\ effectively predicts the initial evolution
of razor-thin self-gravitating discs, and to consider the matter settled.
However, a closer examination reveals significant discrepancies between the results of \citetalias{Sellwood2012}, \citetalias{Fouvry+2015}, and our findings.
Specifically, the timing of linear instability in \fig~\ref{fig:discs_fluctuations} and the shape of the flux in \fig~\ref{fig:discs_BL_NBODY} deserve further scrutiny.
Indeed, let us point out that the ensemble-averaged predictions as well as measurements from \fig~\ref{fig:discs_BL_NBODY} present rather broad signatures in action space.
These differ from the sharp ridge-like features presented in \citetalias{Sellwood2012}'s \Nbody\@ simulation or in \citetalias{Fouvry+2015}'s \BL\@ prediction.
Furthermore, the overall amplitude of the flux is yet to be convincingly explained.

To explore these issues, let us now address the following questions:
(i) what is the long-term impact of collective effects?
(ii) What role do (weakly) damped modes play in this long-term evolution?
(iii) Does softening have any long-term signatures?
(iv) How similar is an individual disc to the ensemble average?
We investigate these questions in order.

\subsection{Neglecting collective effects}
\label{sec:Landau}

\begin{figure}
\begin{center}
\includegraphics[width=0.48\textwidth]{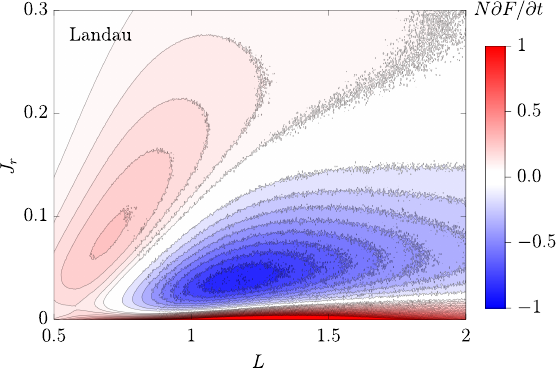}
\caption{Relaxation rate, $\p F/\p t$, as predicted by the Landau equation (see \app~\ref{app:Landau}),
using the same convention as in \fig~\ref{fig:discs_BL_NBODY}. The Landau equation predicts a more isotropic diffusion.
In addition, the relaxation time predicted by \BL\ is three orders of magnitude shorter than the one predicted by Landau:
collective amplification is instrumental.
}
\label{fig:discs_longterm_predictions_Landau}
\end{center}
\end{figure}

In dynamically hot systems,
i.e., systems with large velocity dispersions,
collective effects can be neglected.
Then, \BL\@ reduces to the Landau equation~\citep[see][and references therein]{Chavanis2013stellar}.
As detailed in \app~\ref{app:Landau},
this amounts to replacing in \eq~\eqref{eq:BL}
the dressed susceptibility coefficients, $U^{\rd}_{\bk\bk'}$,
with their bare counterpart, $U_{\bk\bk'}$.
Figure~\ref{fig:discs_longterm_predictions_Landau} 
presents the relaxation rate in action space,
as predicted by the inhomogeneous Landau equation.
There are two main takeaways from this figure.

First, as already pointed out by \citetalias{Fouvry+2015},
the \BL\ relaxation rate is larger than the Landau one by three orders of magnitude.
Phrased differently, collective effects considerably accelerate the relaxation.
In the present half-mass Mestel disc (${ Q \!=\! 1.5 }$),
\cite{Toomre1981} has shown that collective effects can swing amplify~\citep{GoldreichLyndenBell1965b,JulianToomre1966}
perturbations by a factor ${|U^{\rd}_{\bk\bk'} / U_{\bk\bk'}| \!\simeq\! 30}$ (\extfig~{7} therein).
Since \BL\ involves the dressing squared,
i.e., ${ |U^{\rd}_{\bk\bk'}|^{2} }$,
this leads to a considerable acceleration of the relaxation.
One gets ${|\p_{t} F^{\mathrm{BL}}|/|\p_{t} F^{\mathrm{Landau}} | \!\simeq\! |U^{\rd}_{\bk\bk'} / U_{\bk\bk'}|^2 \!\simeq\! 10^3 }$.

Second, we note that the \BL\@ relaxation rate (\fig~\ref{fig:discs_BL_NBODY})
presents much sharper structures in action space compared
to the Landau one (\fig~\ref{fig:discs_longterm_predictions_Landau}).
Yet, both kinetic equations are inhomogeneous
and account for resonant encounters.
We argue that the narrower features predicted by \BL\
stem from the imprint of the disc's weakly damped modes.
These modes drive strong localised responses in the disc.
This is what we explore in the next section.

\subsection{Weakly damped modes}
\label{sec:longterm_damped_modes}

\begin{figure}
\begin{center}
\includegraphics[width=0.48\textwidth]{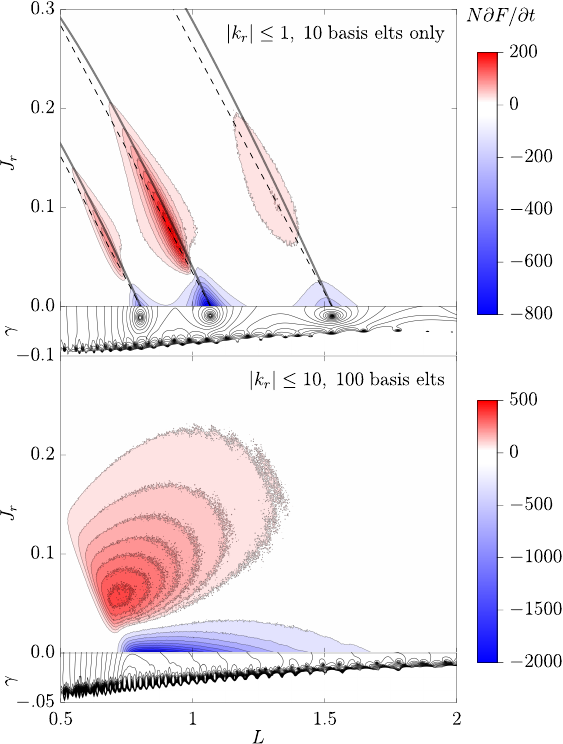}
\caption{Illustration of the connexion between a disc's long-term relaxation
and the disc's linear damped modes.
In each panel, we represent in the top plot the relaxation rate in action space, predicted by \BL\@, using the same convention as in \fig~\ref{fig:discs_BL_NBODY}.
In the bottom plot, we represent the disc's susceptibility, ${ |\mathbf{N} (\omega)| }$ (\eq~\ref{eq:dressed_coupling_coefficients_effective}) represented at the circular angular momentum corresponding to their \ILR\@  (see \eq~\ref{eq:translation_ILR}).
The numerical saturation of the susceptibility for damped frequencies, ${\gamma\!<\!0}$, is to be expected \citep[see, e.g., Appendix~{B} in][]{Petersen+2024}.
\textit{Top}:~Prediction purposely using too small a number of basis elements (\app~\ref{app:bases}) and too few resonances, for illustration purposes.
In that case, the disc's susceptibility contains three clear damped modes.
These modes have a direct signature in the \BL\ relaxation rate.
The black line corresponds to the \ILR\ resonance line associated with each mode  while the dashed lines correspond to the direction of diffusion. The proximity of these two lines enhances the efficiency of \ILR\ for heating the disc.
\textit{Bottom}:~Same as above, but using a numerically converged linear susceptibility,
namely using the same parameters as in \fig~\ref{fig:discs_BL_NBODY}.
In both figures, we show that the disc's long-term heating is strongly enhanced at resonance with the underlying weakly damped modes.
}
\label{fig:ridges_modes}
\end{center}
\end{figure}

To understand the role played by (weakly) damped modes we illustrate, in \fig~\ref{fig:ridges_modes}, the \BL\ relaxation rate on top of the disc's linear susceptibility.
Figure~\ref{fig:ridges_modes} is an intricate figure,
that we now describe carefully.

Figure~\ref{fig:ridges_modes} is composed of two panels, each of them with two sub-plots.
In both panel, the top plot is the \BL\ relaxation rate in action space,
using the same convention as in \fig~\ref{fig:discs_BL_NBODY}.
The bottom plot represents the disc's susceptibility,
through the determinant of the susceptibility matrix, ${ |\mathbf{N} (\omega) |}$
(\eq~\ref{eq:dressed_coupling_coefficients_effective}).
A mode corresponds to a (complex) frequency, $\omega$,
such that ${ |\mathbf{N} (\omega)| }$ diverges,
i.e., they appear as poles in this panel.
Yet, here, we do not represent the susceptibility in frequency space,
but rather translate ${ \RePart [\omega] }$ into an action
through the circular angular momentum of the associated \ILR\@.
More precisely, to any ${ \RePart[\omega] }$,
we can unambiguously associate
an angular momentum, $L$, by solving
\begin{equation}
\bk_{\mathrm{ILR}} \!\cdot\! \bOmega (J_{r} \!=\! 0 , L) = \RePart[\omega] ,
\label{eq:translation_ILR}
\end{equation}
with the resonance vector ${ \bk_{\mathrm{ILR}} \!=\! (k_{\mathrm{ILR}}^{r} , k_{\mathrm{ILR}}^{\phi}) \!=\! (-1,2) }$.
With this convention, if a (quasi-circular) orbit resonates with the mode through its \ILR\@,
then the location of the orbit in action space will lie just above
the associated mode in the susceptibility plot.

In \fig~\ref{fig:ridges_modes}, the two panels
correspond to the same disc model
but use different calculations of the linear susceptibility.
The top panel uses, purposely, a reduced number of basis elements (\app~\ref{app:bases}),
hence degrading the convergence of the linear predictions.
This allows for the physical processes at play here to appear more clearly.
The bottom panel uses the same parameters as in \fig~\ref{fig:discs_BL_NBODY},
with a much larger number of basis elements.
It therefore relies on a converged estimation of the linear susceptibility.

Let us now focus on the top panel of \fig~\ref{fig:ridges_modes}.
In that degraded case, the linear calculation
predicts (at least) three weakly damped modes.
In the associated \BL\ prediction, one notes that each of these modes
is at the origin of a sharp resonant ridge in action space.
More precisely, the real part of the mode's frequency
dictates the position of the induced ridge in action:
the larger the mode's pattern speed,
the further in the resonant ridge appears.
The imaginary part of the mode's frequency dictates the width
of the induced ridge in action space:
the faster the mode's damping,
the wider the imprint of the ridge in action space.
Overall, we note that the resonant signatures of the damped modes,
because they drive such a strong collective amplification,
fully dominate the long-term relaxation predicted by \BL\@.

The same mechanism is operating in the bottom panel of \fig~\ref{fig:ridges_modes}.
In this case, the disc's linear susceptibility along the real axis is large within a broad frequency region, and does not display sharp signatures of distinguishable damped modes (given the shapes of the isocontours).
This, in turn, drives significant amplification across a wide range of angular momenta, resulting in the broader heating in action space predicted by \BL\@.
As a consequence, \BL\ does not exhibit distinct ridges in action space.
Naturally, it would be highly valuable to refine our current implementation of linear response theory to overcome the numerical saturation evident in \fig~\ref{fig:ridges_modes} and gain a clearer understanding of the modal structure of the Mestel disc.

Orbits that resonate with a mode through their \ILR\@
flow in action space along the direction of $\bk_{\mathrm{ILR}}$.
Phrased differently, these orbits
are heated toward more eccentric orbits
with smaller guiding radii.
It is worth noting that, coincidentally for the \ILR\@, this direction of diffusion aligns closely with the associated resonance line, i.e., the line of constant ${ \bk_{\mathrm{ILR}} \!\cdot\! \bOmega (\bJ) }$ (black line in the top panel of Fig.~\ref{fig:ridges_modes}).
This alignment makes the \ILR\ a particularly efficient resonance for heating the disc.

\begin{figure}
\begin{center}
\includegraphics[width=0.4\textwidth]{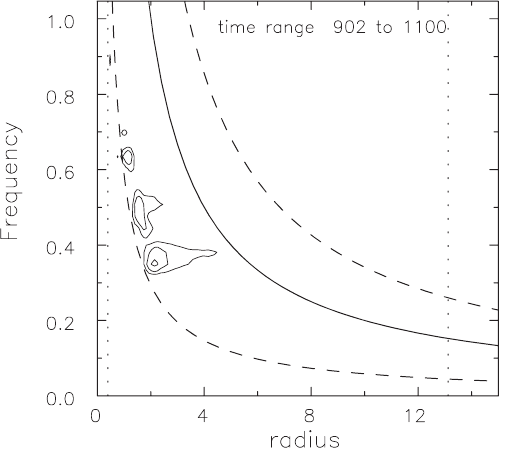}
\caption{Figure from \protect\citetalias{Sellwood2012}.
Power-spectra of the ${\ell\!=\!2}$ fluctuations as a function of radius in their \Nbody\ simulation.
The presence of three peaks in these early power-spectra,
and their location at their respective \ILR\@,
is likely the resonant signature of weakly damped modes.
}
\label{fig:sellwood12_modes}
\end{center}
\end{figure}

Interestingly, \citetalias{Sellwood2012} observed the local power spectrum of fluctuations
in their \Nbody\ simulations, before any instability has kicked in.
Their resulting~\extfig~{4} is reproduced here in \fig~\ref{fig:sellwood12_modes}.
In that figure, three peaks can be observed in the power spectrum, each corresponding to its respective \ILR\@ radius.
These peaks are possibly associated with weakly damped modes excited by Poisson shot noise.
While their number appears tentatively comparable to our results, any such comparison should be approached with caution.
Indeed, the prediction in the top panel of \fig~\ref{fig:ridges_modes} was degraded
for illustration purposes.
Obviously, improving the convergence of the linear predictions
is necessary for more quantitative comparisons.
In addition, one could also experiment with the method from~\cite{Weinberg1994,Heggie+2020}
to track damped modes in \Nbody\@ simulations.

Finally, let us briefly comment on the difference between the \BL\ prediction from~\citetalias{Fouvry+2015} (\extfig~{4} therein) and our current \BL\ prediction (\fig~\ref{fig:discs_BL_NBODY}).
In \citetalias{Fouvry+2015}, \BL\ predicted a single sharp ridge in action space, whereas, in the present work, \BL\ predicts broader heating in action space.
Since the \BL\@'s prediction is closely tied to the accuracy of the linear susceptibility, the discrepancy between these two studies stems from the computation of the dressed coupling coefficients, $U^{\rd}_{\bk \bk'}$, which were badly converged in~\citetalias{Fouvry+2015}.
Appendix~\ref{app:linear-response} details the improved methodology used in the present work to compute
these coefficients more accurately.

\subsection{Impact of softening}
\label{sec:longterm_discs_softening}

In \app~\ref{app:discs_stability_softening},
following~\cite{DeRijcke+2019b},
we investigate how softening affects the complex frequency of growing modes in an unstable disc.
In particular, in \fig~\ref{fig:DeRijcke_modes},
we show that Plummer softening (\eq~\ref{eq:Plummer_softening}) tends to shift unstable modes
toward smaller pattern speed and growth rates.
As a consequence, if the same trend also applies to damped modes,
we expect that the stronger the softening, the faster the damping of the modes.
Rephrased using the insight from \fig~\ref{fig:ridges_modes},
we expect that the stronger the softening,
(i) the later the action space ridges should appear,
(ii) the more these ridges would move toward higher angular momenta.
We set out to investigate these trends in \Nbody\ simulations.

\begin{figure}
\begin{center}
\includegraphics[width=0.48\textwidth]{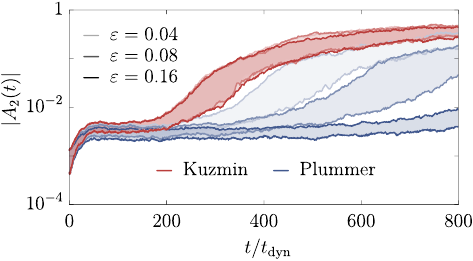}
\caption{Same as \fig~\ref{fig:discs_fluctuations}
but for various softening lengths, $\veps$, and two different softening kernels (\app~\ref{app:discs_stability_softening}).
The level of transparency correspond to different softening lengths (\eq~\ref{eq:softening_kernel})
with the bottom (resp.\ top) line corresponding to 20\% (resp.\ 80\%)
contours over 100 independent realisations with ${ N \!=\! 25 \!\times\! 10^{6} }$ particles.
The case `Kuzmin ${ \veps \!=\! 0.16 }$' corresponds to \fig~\ref{fig:discs_fluctuations}.
For the Plummer kernel, the larger the softening length, the slower the relaxation, and the more delayed the transition to instability.
This kernel introduces a strong gravitational bias (\fig~\ref{fig:DeRijcke_modes}).
On the contrary, simulations using the Kuzmin kernel behave similarly across softening lengths.
Interestingly, we note that, in every case, the dispersion among realisations increases near phase transition.
}
\label{fig:longterm_softening}
\end{center}
\end{figure}

In \fig~\ref{fig:longterm_softening}, we illustrate the long-term evolution of bisymmetric fluctuations
in \Nbody\ simulations for the two softening kernels presented in \app~\ref{app:discs_stability_softening} with different values of the softening length, $\veps$.
In that figure, it is clear that, for the Plummer softening kernel, increasing $\veps$
delays the phase transition toward instability.
We claim that this is so because softening affects the collisionless properties of the disc,
and consequently its collisional dressed resonant relaxation.
Phrased differently, (Plummer) softening makes the disc more linearly stable than it should be.
Consequently, softening reduces swing amplification and ultimately delays relaxation.\footnote{The same trend with softening would also happen if the disc's relaxation was driven
by two-body local encounters~\citep{Theis1998}.
However, such local contributions are absent from our experiments,
since our simulations are, by design, limited to ${\ell\!=\!2}$ fluctuations.}

The trend from \fig~\ref{fig:longterm_softening} also explains
the difference in the time of phase transition between \citetalias{Sellwood2012}'s simulation and ours.
Indeed, in \extfig~{2} of~\citetalias{Sellwood2012},
the instability sets in at ${ t \!\simeq\! 1\,200 \, \tdyn }$
using the Plummer softening with ${ \veps \!=\! 0.125 }$ and ${ N \!=\! 50 \!\times\! 10^{6} }$.
In our case (\fig~\ref{fig:longterm_softening}), using ${ N \!=\! 25 \!\times\! 10^{6} }$,
the instability sets in at ${ t \!\simeq\! 400 \, \tdyn }$ (resp.\ $800$)
for the Plummer softening with ${ \veps \!\simeq\! 0.08 }$ (resp.\ ${ 0.16 }$).
Once accounting for the rescaling by the factor ${1/N}$,
these timescales are nicely consistent.

We conclude this section by stressing
the importance of accounting for the bias introduced by softening
when comparing \Nbody\ simulations with both linear response and kinetic theory. 
In \fig~\ref{fig:discs_BL_NBODY}, we carefully chose a softening kernel, namely Kuzmin softening (\eq~\ref{eq:Kuzmin_softening}), so that the corresponding bias is small.
Indeed, this kernel affects much less the disc's linear response (see \fig~\ref{fig:DeRijcke_modes}),
so that varying the softening length does not bias their long-term relaxation (\fig~\ref{fig:longterm_softening}).

\subsection{Long-term stochasticity}
\label{sec:stochastic}

\begin{figure}
\begin{center}
\includegraphics[width=0.48\textwidth]{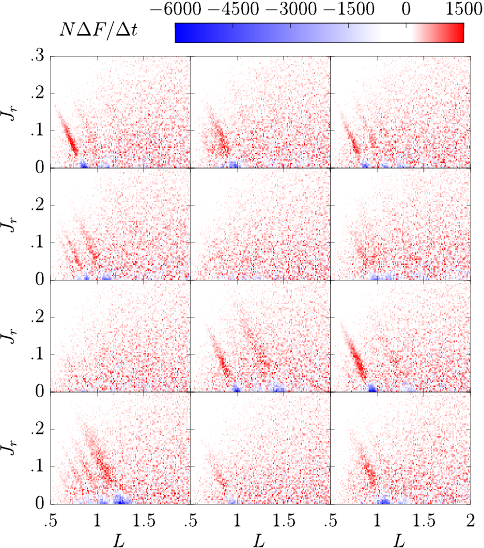}
\caption{Relaxation rate, ${ \p F / \p t }$, measured in twelve independent \Nbody\ realisations,
using the same convention as in \fig~\ref{fig:discs_BL_NBODY}.
The number, location, amplitude and birth time of the action ridges
vary strongly from one realisation to the other.
This is a measure of  a long-term stochasticity in the vicinity of the phase transition (see also \fig~\ref{fig:slices}).
}
\label{fig:longterm_stochasticity}
\end{center}
\end{figure}

\BL\ predicts the ensemble-averaged evolution,
i.e., averaged over different initial conditions drawn from the same \DF\@.
It is also of interest to investigate the dispersion that may exist
among different realisations of the same disc.
To do so, we represent in \fig~\ref{fig:longterm_stochasticity} the relaxation rate measured in twelve
independent simulations whose fluctuations are presented in \fig~\ref{fig:discs_fluctuations}.

From \fig~\ref{fig:longterm_stochasticity}, one can undoubtedly note that individual realisations
are very different from the averaged one (\fig~\ref{fig:discs_BL_NBODY}).
Indeed, on the ensemble-averaged figure,
all the circular orbits (${ 0.75 \!\lesssim\! L \!\lesssim\! 1.5 }$)
are depopulated in profit of more eccentric orbits.
Yet, in individual realisations, the changes in the \DF\
are much more localised,
and concentrated along one or multiple sharp ridges in action space.
These ridges come in different number, locations,
and at different time/amplitude in each realisation.\footnote{We note that \citetalias{Sellwood2012} and~\citetalias{Fouvry+2015} focused on a single realisation and did not perform any ensemble average.}
We also note that their individual intensity is up to 2-3 time larger
than the averaged one.

\BL\ is, by design, unable to predict
such a diversity.
Predicting the scatter among realisations could be addressed within the context of large deviation theory~\citep[see, e.g.\@,][]{FeliachiBouchet2022}.
However, the inclusion of collective effects in this context is still an open question~\citep{FeliachiFouvry2024}.
As such, the details of the processes responsible for this long-term stochasticity are still to be understood.

\section{Conclusion}
\label{sec:conclusion}

We presented applications of the \BL\@ kinetic theory to the self-induced long-term evolution of razor-thin self-gravitating discs.
We showed that the inhomogeneous \BL\@ \eq~\eqref{eq:BL} is able to quantitatively predict the mean long-term evolution of these systems.
The kinetic predictions were succesfully compared with \Nbody\@ simulations
averaged over many realisations (\fig~\ref{fig:discs_BL_NBODY}).
Doing so, we highlighted the importance of resonant encounters and collective effects in driving the relaxation of stellar discs.
The consequences of self-gravity proved to be particularly diverse (\fig~\ref{fig:discs_longterm_predictions_Landau})
and would have been difficult to predict \textit{a priori} without the valuable insights provided by \BL\@.

In these marginally stable systems,
fluctuations get strongly swing amplified by resonating
with the disc's underlying weakly damped modes (\fig~\ref{fig:ridges_modes}).
The disc's long-term evolution is therefore particularly sensitive to the collisionless linear response of the disc.
By stabilizing the disc, we showed that softening delays the phase transition to instability (\fig~\ref{fig:longterm_softening}).
We also showed that the choice of softening kernel was critical:
in razor-thin discs, the Plummer softening should be avoided both for linear and long-term responses to compare with (non-softened) theoretical predictions.
Finally, we noted that, in this regime, the average relaxation rate is not representative of the individual realisations.
As the disc nears marginal stability, the variance among realisations gets increasingly large (\fig~\ref{fig:longterm_stochasticity}).
In that sense, ``the average cold galaxy does not exist''.

Let us conclude with a few prospects worth exploring to further our theoretical understanding of the long-term evolution of self-gravitating discs.

\textit{Small scale vs large scale} -- \BL\ diverges on small scales~\citep[see, e.g.\@,][]{Chavanis2013stellar}.
\cite{Chandrasekhar1943} did regularise the deflections at small scales taking hard encounters into account.
This regularisation should be adapted to the resonant kinetic theories~\citep[see, e.g.\@,][]{Fouvry+2021},
therefore weighing the respective impact of small vs large scales on long-term relaxation.
In (cold) razor-thin galactic discs, the long-range dressing by collective effects is likely to dominate,
but this remains to be accurately quantified.

\noindent\textit{Time evolution} --
A direct follow-up of this work would involve integrating \BL\@ forward in time,
beyond the onset of relaxation.
Yet, integrating these equations is challenging~\citep{Weinberg2001}.
There are two main difficulties to overcome:
(i) computing the resonant diffusion coefficients;
(ii) determining the self-consistent potential (and actions) as the disc evolves.
One potential approach is to integrate the corresponding stochastic Langevin equation~\citep[see, e.g.\@,][]{Henon1971,Cohn1979,Giersz1998,Fu+2024}.
This could ultimately serve to validate \Nbody\@ codes on longer timescales.

\noindent\textit{Crossing marginal stability} --
\BL\@ unrealistically diverges at marginal stability~\citep{Weinberg1993},
though we showed here that \BL\ is still accurate relatively close to the phase transition.
Yet, if one was to evolve \BL\@ in time, it would inevitably break at some point.
This divergence must be regularised by considering mode-particle interactions,
in the spirit of the so-called quasilinear theory in plasma physics \citep[see, e.g.,][]{RogisterOberman1968,HamiltonHeinemann2020,HamiltonHeinemann2023}.

\noindent\textit{Non-linear regime} --
After the phase transition, the disc's evolution is dominated by the emerging unstable mode.
As this mode grows, non-linear contributions become increasingly important,
up to the point of saturation and trapping at resonances~\citep[see, e.g.,][]{Hamilton2024}.
Investigating this late-time non-linear regime is an interesting venue.

\noindent\textit{Beyond average predictions} --
As highlighted in \fig~\ref{fig:longterm_stochasticity},
near marginal stability  the variance among different realisations
increases tremendously.
Predicting this scatter is crucial to assess the likelihood of a given realisation.
This is beyond the reach of \BL\@,
but large deviation theory offers promising perspectives~\citep[see, e.g.\@,][]{Bouchet2020,FeliachiBouchet2022,FeliachiFouvry2024}.
Ultimately, this should pave the way for the comparison of theoretical predictions of the scatter 
with observations and cosmological simulations.

\noindent\textit{Thick discs} --
Of course, the models studied here are not perfect representation of genuine galactic discs.
While the case of spherically symmetric systems is well captured by \BL\@~\citep{Rozier+2019,Fouvry+2021,Tep+2022},
the extension to thick discs or flattened spheres is yet to be achieved
beyond the WKB approximation~\citep{Fouvry+2017}.
One of the potential issue in these systems is the possible lack of integrability~\citep[see, e.g.\@,][]{Weinberg2015}
beyond the St\"ackel family~\citep{Tep+2024b}.

\noindent\textit{Coupled halo-disc evolution} --
We assumed that the disc was embedded within a static, rigid dark matter halo.
Yet, the dark matter halo, albeit dynamically hotter, is also subject to fluctuations.
Accounting for the coupling between the disc and the halo should provide a more realistic picture of the long-term evolution of galaxies \citep[see, e.g.,][]{Johnson+2023}.

\noindent\textit{Impact of morphological type} -- The present analysis focussed on the Mestel disc. It would be of interest to study the efficiency of orbital diffusion  on more realistic exponential disc models~\citep[see, e.g.\@,][]{DeRijcke+2019a}
while  varying, e.g.\@, the bulge to disc mass,
when scanning late galaxy types of the Hubble sequence~\citep{Reddish+2022}.

\noindent\textit{Open dissipative systems} --
Finally, galaxies are not isolated objects.
The Milky Way is for instance continuously perturbed by external sources triggering various response features \citep[see, e.g.,][for a review]{GrionFilho+2021}
that operate concurrently in open systems.
In addition, the accounting of dissipative processes within the baryonic component
can drive unlikely evolutionary pathways,
possibly leading to self-regulation \citep{Pichon2023-bs}.
This should also be the topic of future work.

\subsection*{Data Distribution}
The data underlying this article 
is available through reasonable request to the author.
The code for the diffusion coefficient, written in \julia\@~\citep{JuliaCite},
is available at the following URL: \href{https://github.com/MathieuRoule/SecularResponse.jl}{https://github.com/MathieuRoule/SecularResponse.jl}.
The \Nbody\@ code is also publicly available (courtesy of John Magorrian) at \href{https://github.com/MathieuRoule/mestel2d}{https://github.com/MathieuRoule/mestel2d}.

\section*{Acknowledgements}

This work is partially supported by grant Segal ANR-19-CE31-0017
of the French Agence Nationale de la Recherche,
and by the Idex Sorbonne Universit\'e.
This research was supported in part by grant NSF PHY-2309135 to the Kavli Institute for Theoretical Physics (KITP).
We are grateful to J.~Barr\'{e}, E.~Donghia, S.~Flores, E.~Ko, J.~Magorrian, M.~Petersen, K.~Tep, and M.~Weinberg for constructive comments.
We thank J.~Magorrian for providing us with his code,
and St\'ephane Rouberol for the smooth running of the
Infinity cluster, where the simulations were performed.

\appendix

\section{Razor-thin disc}
\label{app:model_disc}

\subsection{Angle-action coordinates}
\label{app:disc_AA}

Axisymmetric razor-thin discs are integrable,
i.e., appropriate angle-action coordinates can explicitly be constructed for them~\citep{LyndenBellKalnajs1972}.
In practice, we follow the same approach as in the library \orbitalelements\@,
whose main conventions we recall here~\citepalias[see \extapp~{A1} in][]{Petersen+2024}.

Orbits in a razor-thin disc can be equivalently labelled
by their energy and angular momentum, ${ (E,L) }$,
their action ${ (J_{r} , L) }$ -- with $J_{r}$ the radial action,
or with their pericentre and apocentre, ${ (\rper, \rapo) }$~\citepalias[see \exteqs\ {A1}--{A3} in][]{Petersen+2024}.
Orbits are skimmed with frequencies ${ (\Omega_{r} , \Omega_{\phi}) \!=\! (\alpha \Omega_{0} , \alpha \beta \Omega_{0}) }$,
with ${ \Omega_{0} }$ some frequency scale.
The frequency ratios ${ (\alpha , \beta) }$
readily follow from the angle-action mapping~\citepalias[see \exteq~{A4} in][]{Petersen+2024}.
Following \extapp~{A1.1} in~\citetalias{Petersen+2024},
all angular integrals are performed using the so-called ``H\'enon anomaly'',
with additional interpolations to handle extreme orbits,
e.g.\@, exactly circular or radial orbits.

\subsection{Distribution function}
\label{app:disc_DF}

The \DF\ of the Mestel disc from \eq~\eqref{eq:Mestel_DF}
satisfies~\citep[\exteq~{4.163} in][]{BinneyTremaine2008}
\begin{equation}
\label{eq:Mestel_DF_constants}
q \!=\! \bigg( \frac{V_0}{\sigma} \bigg)^2 \!\!-\! 1 , \,\,
C \!=\! \frac{V_0^2}{2^{q/2+1}\pi^{3/2}\,G\;\Gamma[\tfrac{q+1}{2}] \sigma^{q+2}R_0^{q+1}} .
\end{equation}
In order to deal with the central singularity and the disc's infinite mass, we follow~\cite{EvansRead1998}
and introduce an inner and outer tapering in the \DF\@,
leaving the mean potential unchanged.
The  total potential is then generated by
(i) an inert bulge, 
(ii) an inert halo and (iii) the self-gravitating disc.
In addition, one may vary the overall amplitude of the \DF\@ with   the active fraction $\xi$.
This acts as a proxy for the relative masses of the disc and its surrounding halo.
The disc's \DF\@ reads
\begin{equation}
\label{eq:tapered_Mestel_DF}
F(E,L) = \xi \, C \, L^q \, \re^{-E/\sigma^2} \,
\Tin(L) \, \Tout(L).
\end{equation}
In this equation, the tapers read
\begin{equation}
\label{eq:tapers_Mestel}
\Tin (L) \!=\! \frac{L^{\nu}}{(\Rin V_0)^{\nu}+L^{\nu}}, \,\,\, 
\Tout (L) \!=\! \frac{(\Rout V_0)^{\mu}}{(\Rout V_0)^{\mu}+L^{\mu}}.
\end{equation}
The sharpness of the inner (resp.\ outer) taper is controlled by the power index $\nu$ (resp.\ $\mu$) while its location is set by the radius $\Rin$ (resp.\ $\Rout$).
The outer taper was introduced by~\cite{EvansRead1998},
 and is mandatory in \Nbody\@ simulations
to ensure that the disc is of finite mass.

The exact set of parameters we used are the same as in~\citetalias{Sellwood2012} and~\citetalias{Fouvry+2015}.
More precisely, in \fig~\ref{fig:discs_BL_NBODY},
we take ${ G \!=\! R_0 \!=\! V_0 \!=\! 1 }$
and use these units throughout the paper. 
We also imposed ${ \xi \!=\! 0.5}$ (half-mass disc),
${ \eps \!=\! 10^{-5} }$ for the potential truncation (see footnote~\ref{fn:soft_pot}),
and fixed the tapers to ${ \Rin \!=\! 1}$, ${\Rout \!=\! 11.5}$,
${\mu \!=\! 5}$, and ${ \nu \!=\! 4}$.
Finally, we set the disc's velocity dispersion to ${q \!=\! 11.4}$.
The value of $q$ slightly differs between
\citetalias{Sellwood2012} (${ q \!=\! 11.44}$) and \citetalias{Fouvry+2015} (${ q \!=\! 11.4}$).
In practice, this only changes the velocity dispersion by ${ 0.1\% }$ and we checked that it did not impact our predictions.
\citetalias{Sellwood2012}'s value had a simple motivation: having nominal Toomre's factor ${Q\!=\!1.5}$.

\section{\Nbody\ simulations}
\label{app:Nbody}

To perform our \Nbody\@ simulation, we adapted the particle-mesh code used in \citetalias{Fouvry+2015} (courtesy of John Magorrian), a simpler 2D version of the {\tiny\texttt{GROMMET}} code \citep{Magorrian2007}.
We refer to \extsect~{5.1} of \citetalias{Fouvry+2015} for a detailed description of the code.
The \Nbody\@ code is publicly available (courtesy of John Magorrian) at \href{https://github.com/MathieuRoule/mestel2d}{https://github.com/MathieuRoule/mestel2d}.

\subsection{Sampling}
\label{app:disc_sampling}

Improving upon \extapp~{E} of \citetalias{Fouvry+2015},
we sample the DF from \eqref{eq:tapered_Mestel_DF}
in ${ (\rper,\rapo) }$--space rather than in ${ (E,L) }$--space.
Let us briefly detail our approach.

Following \eq~\eqref{eq:tapered_Mestel_DF},
we need to sample the \DF\
\begin{equation}
F_{\mathrm{sp}}(E,L) = C_{\mathrm{sp}} \, L^q \, \re^{-E/\sigma_r^2} \, \Tin(L) \,\Tout(L) ,
\label{eq:Zang_sampling_DF}
\end{equation}
complemented with the truncation constraint\footnote{As in \citetalias{Fouvry+2015}, we interpret this constraint as ``no particles with orbits that extend beyond $\Rmax$'', i.e., ${F(E,L)\!=\!0}$ when ${E\!>\!\psieff(\Rmax)}$ [and not $\psi(\Rmax)$, see~\citetalias{Sellwood2012}].} 
\begin{equation}
\label{eq:Rmax_truncation}
\rapo(E,L) \leq \Rmax .
\end{equation}
In \eq~\eqref{eq:Zang_sampling_DF},
the constant ${ C_{\mathrm{sp}} }$ ensures that $F_{\mathrm{sp}}$
is normalised to unity when integrated w.r.t.\ ${ \rd \bx \rd \bv }$.
We report the values used in \tab~\ref{tab:Zang_normalisation}.
\begin{table}
\tiny
\centering
\begin{tabular}{| C{0.18\columnwidth} | C{0.23\columnwidth} | C{0.17\columnwidth} | C{0.19\columnwidth} |}
\hline
Distribution function & Normalisation ${ C_{\mathrm{sp}} / C }$ & Total mass $\Mtot$ & Rejection constant $M$\\
\hline\hline
Mestel & $\simeq 9.25\!\times\!10^{-2}$ & $10.8\,(\times\xi)$ & 15.7 \\
\hline
Zang $\nu\!=\!4$ & $\simeq 9.33\!\times\!10^{-2}$ & $10.7\,(\times\xi)$ & 11.8 \\
\hline
\end{tabular}
\caption{Constants used for the sampling of the \DFs\@ from \eq~\eqref{eq:Zang_sampling_DF}.
The total mass $\Mtot$ is not set to one and corresponds to the mass enclosed within the truncation radius $\Rmax$ (\eq~\ref{eq:Rmax_truncation}).
The individual mass of the particles is set to ${m\!=\!\Mtot / N}$.
The (minimal) rejection constants correspond to the maximal value of the density of state in ${ (\rper, \rapo) }$
from \eq~\eqref{eq:Zang_density_of_state}.
}
\label{tab:Zang_normalisation} 
\end{table}

The constraint from \eq~\eqref{eq:Rmax_truncation}
imposes that the populated domain in the ${ (\rper,\rapo) }$--space
is a triangle. Accounting carefully for the Jacobian of the transformation
${ (J_{r} , L) \!\to\! (\rper, \rapo) }$, the density of state in ${ (\rper,\rapo) }$ is
\begin{equation}
p(\rper,\rapo) = \frac{(2\pi)^{2}}{\Omega_r}
\left| \frac{\p (E,L)}{\p (\rper,\rapo)} \right|
F_{\mathrm{sp}}(E,L),
\label{eq:Zang_density_of_state}
\end{equation}
where the ${ (2 \pi)^{2} }$ factor comes from integrating over the angles,
and the Jacobian ${ |\p (E,L) / \p (\rper,\rapo)| }$
is computed using \orbitalelements\@~\citepalias{Petersen+2024}.

To sample ${ p (\rper,\rapo) }$, we
use a rejection sampling against the uniform distribution
\begin{equation*}
g(\rper,\rapo) = \frac{2}{\Rmax^2}\mathds{1}_{\{0\leq\rper\}} \, \mathds{1}_{\{\rper\leq\rapo\}}
\, \mathds{1}_{\{\rapo\leq\Rmax\}} ,
\end{equation*}
with the rejection constant $M$ satisfying
\begin{equation}
\forall (\rper,\rapo), \; p(\rper,\rapo) \leq \frac{2M}{\Rmax^2} .
\end{equation}
The values used for this rejection constant are given in \tab~\ref{tab:Zang_normalisation}.

\subsection{Measuring fluctuations and modes}
\label{app:Measuring_modes}

Let us now describe our method to measure the
fluctuation's strength (\figs~\ref{fig:discs_fluctuations} and \ref{fig:longterm_softening})
and to estimate the frequency and growth rate of the dominant mode in \Nbody\@ simulations of unstable discs (\app~\ref{app:discs_stability_softening}).
We follow an approach similar to \extapp~{C} in~\citetalias{Fouvry+2015}.

\begin{figure}
\centering
\includegraphics[width=0.48\textwidth]{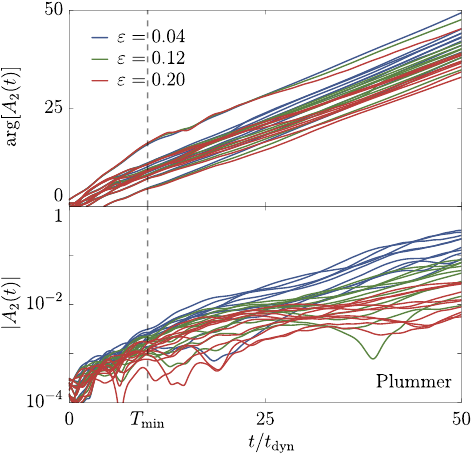}
\caption{Time series of the unwrapped phase (top) and the norm (bottom) of the bisymmetric fluctuations probed by $A_2(t)$ from \eq~\eqref{eq:mode_projection} over 10 independent realisations of the unstable Zang ${\nu\!=\!4}$ disc \citep{ZangThesis}.
Each simulation is performed with ${N\!=\!2\!\times\!10^{8}}$ particles using the Plummer softening kernel (\eq~\ref{eq:Plummer_softening}) and for varying softening lengths, $\veps$.
For each $\veps$, the same ten initial conditions are used and shown in the same colour.
The frequency $\OmegaM$ (resp.\ growth rate $\gammaM$) of the mode are estimated by a linear fit of the phase (resp.\ log of the norm) of such time series, only using ${t \!\geq\! T_{\min} \!=\! 10 \, \tdyn }$.
The growth of the phase is generically more regular than the growth of the norm.
The Plummer softening kernel introduces a strong gravity bias.
}
\label{fig:modes_time_series}
\end{figure}

Considering that the system's density fluctuations are well described by a single dominant mode, they read 
\begin{equation}
\label{eq:mode_density}
\delta\rho(\br,t) = \rho_{\scM}(\br) \exp(\ri \omegaM t),
\end{equation}
with ${\rho_{\scM}}$ the shape of the mode and ${\omegaM}$ its (complex) frequency. 
The density fluctuation can be projected on any spatial function, ${ f(\br) \!=\! f(r) \, \re^{\ri\ell\phi} }$,
to give
\begin{equation}
\label{eq:mode_projection}
A_\ell(t) = \!\!\int\!\! \rd \br f(\br) \, \delta\rho(\br,t) = m \sum_{i = 1}^{N} f[r_i(t)] \, \re^{\ri\ell\phi_i(t)},
\end{equation}
From the time series ${ t \!\to\! A_{\ell} (t) }$,
one expects then the behaviour
\begin{equation}
\label{eq:mode_projection_measure}
|A_\ell(t)| \propto \re^{\gammaM t}; 
\quad \arg[A_\ell(t)] \propto \OmegaM t ,
\end{equation}
provided one ``unwraps'' the phases.
In \fig~\ref{fig:modes_time_series}, we illustrate such time series for an unstable Mestel disc \citep[namely, Zang's ${\nu\!=\!4}$ disc,][]{ZangThesis}.
To measure the dominant mode's frequency,
we use the log-normal function 
\begin{equation}
\label{eq:mode_projection_function}
f(r) = \exp\!\left[-\frac{\ln(r/R_0)^2}{2\sigma^2}\right],
\end{equation}
with ${r_0\!=\!1.33}$ and ${\sigma\!=\!0.45}$.
This mimics the radial shape of the mode,
hence easing the measurement.
If one uses multiple functions for the projections,
one could also estimate the mode's shape.

To investigate long-term evolution of fluctuations in \figs~\ref{fig:discs_fluctuations} and~\ref{fig:longterm_softening},
we use the simpler identity function
\begin{equation}
f(r) = \mathds{1} (r_{\min} \!<\! r \!<\! r_{\max}) ,
\label{eq:mode_projection_longterm}
\end{equation}
with ${ (r_{\min} , r_{\max}) \!=\! (1.5 \, r_{0} , 4.5 \, r_{0}) }$.
This removes the contributions
from the disc's central and outer regions, where very few particles are present.

\subsection{Measuring the relaxation rate}
\label{app:simulations_flux}

To measure the \BL\ relaxation rate in numerical simulations (\fig~\ref{fig:discs_BL_NBODY}),
we performed 1\,000 realisations of the Mestel disc, each with ${N\!=\!25\!\times\!10^6}$ particles.
We used the Kuzmin softening (\eq~\ref{eq:Kuzmin_softening}) with ${ \veps \!=\! 0.16 }$ sufficiently small so that the gravitational bias is negligible (see \fig~\ref{fig:DeRijcke_modes}).
For the numerical integration, we used a leapfrog scheme with timestep ${ \delta t \!=\! 10^{-2} \, \tdyn }$.
The forces are estimated using a Cartesian grid that extends to ${\pm x_{\max}\!=\!20}$ with ${n_x\!\times\!n_y\!=\! 1\,024^2 }$ cells.
The ${\ell\!=\!2}$ fluctuations are selected using the same method as in \extsect~{5.1} of~\citetalias{Fouvry+2015}, with a polar grid of ${n_{\rm r} \!=\! 8\,192}$ radial rings and ${n_{\phi} \!=\! 2\,048}$ points in the azimuthal direction.

For post-processing, we use \orbitalelements\ to compute the particles' actions ${(J_r,L)}$
within the mean potential from \eq~\eqref{eq:Mestel_potential}.
We count their number, ${n(J_i,L_j)}$, in bins of width ${\delta J_r \!=\! 1/300}$, ${\delta L \!=\! 1/100}$.
From these bin counts, the \DF\@ and its changes are obtained through
\begin{align}
\nonumber
n(J_i,L_j) &\simeq (2 \pi)^{2} \frac{N}{\Mtot} \, \delta J_r \, \delta L \, F(J_i,L_j),
\end{align}
where the ${ (2 \pi)^{2} }$ prefactor comes from the integration over ${ \rd \theta_{r} \rd \theta_{\phi} }$.
To estimate ${ \p F / \p t }$ in \fig~\ref{fig:discs_BL_NBODY},
we computed the difference between ${ t \!=\! 150 \, \tdyn }$ and ${ t \!=\! 0}$.

\section{Linear response theory}
\label{app:linear}

As shown in \eq~\eqref{eq:BL}, collective effects play a critical role in shaping the long-term relaxation of dynamically cold systems.
Therefore, it is essential to handle the numerical aspects of linear response theory with precision.
To do so, we use the publicly available library \linearresponse\@~\citepalias{Petersen+2024} and its associated dependencies.
In this section, we briefly outline the main building blocks of this calculation.

\subsection{Biorthogonal bases}
\label{app:bases}

Following the convention from \astrobasis\@~\citepalias{Petersen+2024},
we introduce a basis of potential-density pairs
\begin{subequations}
\begin{align}
\label{eq:basis_int}
{} & \psi^{(p)}(\bw) = \!\! \int \!\! \rd \bw' \, \rho^{(p)}(\bw')\, U(\bw,\bw')
\\
{} & \int \!\! \rd \bw \, \psi^{(p)}(\bw) \, \rho^{(q)*}(\bw) = -\delta_{pq} ,
\label{eq:basis_orth}
\end{align}
\label{eq:def_basis}\end{subequations}
with ${ U \!=\! - G / |\br \!-\! \br'| }$ the gravitational potential.
From these basis elements, the interaction potential takes the pseudo-separable form \citep{HernquistOstriker1992}
\begin{equation}
\label{eq:interaction_potential_basis_split}
U (\bw,\bw') = - \sum_{p} \psi^{(p)}(\bw) \, \psi^{(p)*}(\bw').
\end{equation}

In the present case, the disc's axisymmetry entices us
to introduce basis elements of the form 
${\psi^{(p)} (r,\phi) \!=\! U^{\ell}_{p} (r) \, \re^{\ri\ell\phi} }$ 
with ${U^{\ell}_{p} (r)\!\in\!\mathbb{R}}$.
In practice, we use the radial basis elements from~\cite{CluttonBrock1972}
available in \astrobasis\@.
Advantageously, this basis is
(i) global (one only has to set one scale radius);
(ii) has infinite extent;
(iii) can be computed via a numerically-stable recurrence relation.

\subsection{Response matrix}
\label{app:RepMat}

Linear response is generically captured by the response matrix
${ \bM (\omega) }$~\citep[see, e.g., \exteq~{5.94} in][]{BinneyTremaine2008}
whose elements generically read
\begin{equation}
\label{eq:polarisation_matrix}
M_{pq} (\omega) \!=\! - (2\pi)^{d} \! \!\sum_{\bk \in \mathbb{Z}^{d}} \! \int \!\! \rd \bJ 
\frac{\bk\!\cdot\!\p F / \p \bJ}{\bk\!\cdot\!\bOmega(\bJ) \!-\! \omega} 
\psi^{(p)*}_{\bk} \!(\bJ) \psi^{(q)}_{\bk} (\bJ) ,
\end{equation}
with the Fourier transformed basis elements
\begin{equation}
\psi^{(p)}_{\bk}(\bJ) = \frac{1}{(2\pi)^d} \!\! \int \! \rd\mybtheta \,
\re^{-\ri \bk\cdot\mybtheta} \psi^{(p)}(\bw),
\label{eq:basis_Fourier}
\end{equation}
The definition from \eq~\eqref{eq:polarisation_matrix} holds as such only for ${\ImPart(\omega)\!>\!0}$.
It must be analytically continued to the rest of the complex plane~\citep[see, e.g.\@,][]{FouvryPrunet2022}.

From the response matrix, one defines the susceptibility matrix,
${\bN\!=\![\bI\!-\!\bM(\omega)]^{-1}}$.
The \textit{dressed} coupling coefficients then read~\citep[see, e.g., \exteq~{35} in][]{Heyvaerts2010}
\begin{equation}
\label{eq:dressed_coupling_coefficients_effective}
U^{\rd}_{\bk \bk'}(\bJ,\bJ',\omega) =
-\sum_{p,q} \psi^{(p)}_{\bk}(\bJ) \,
N_{pq}(\omega) \,
\psi^{(q)*}_{\bk'}(\bJ').
\end{equation}
These coefficients describe how stellar orbits
collectively interact with one another.

\subsection{Computations parameters}
\label{app:lr_params}

To compute the response matrix used in \fig~\ref{fig:discs_BL_NBODY}, we used the \linearresponse\ library with the same parameters as \citetalias{Petersen+2024} (see \exttabs~{F1} therein)
which checked for convergence using the unstable Zang ${\nu\!=\!4}$ disc.
Importantly, except when stated differently, we used 100 basis elements and summed over 21 resonances (i.e.\ for ${ |k_{r}| \!\leq\! 10 }$ in \eq~\ref{eq:polarisation_matrix}).
In the top panel of \fig~\ref{fig:ridges_modes}, the same parameters are used,
except for the (specified) number of basis elements and resonances.

\section{Balescu--Lenard equation}
\label{app:BL}

The Balescu--Lenard equation (\eq~\ref{eq:BL}) is a diffusion equation in action space.
Hence, it can be rewritten as
${ \p F (\bJ , t) / \p t \!=\! - \p / \p \bJ \!\cdot\! \flux (\bJ) }$.
The associated flux, ${ \flux (\bJ) }$, is driven by resonances
so that ${ \flux(\bJ) \!=\! \sum_{\bk,\bk'} \! \flux_{\bk\bk'} (\bJ) }$ with
\begin{equation}
\label{eq:BL_single_resonance_generic}
\flux_{\bk\bk'}(\bJ) = \!\! \int \!\! \rd\bJ' \,
G_{\bk\bk'} (\bJ , \bJ') \, \dirac [f_{\bk\bk'}(\bJ,\bJ')] ,
\end{equation}
Here, we introduced the integrand
\begin{align}
\label{eq:BL_G_Flux}
G_{\bk\bk'}(\bJ,\bJ') = &\pi (2\pi)^d \, m \, \bk \, 
\big|U^{\rd}_{\bk\bk'} (\bJ, \bJ', \bk\!\cdot\!\bOmega) \big|^{2}
\nonumber
\\
&\times \bigg[ \bk'\!\cdot\!\frac{\p F}{\p \bJ'}\,F(\bJ) \!-\! \bk\!\cdot\!\frac{\p F}{\p \bJ}\,F(\bJ') \bigg] .
\end{align}
and the resonance condition ${f_{\bk\bk'}(\bJ , \bJ')\!=\! \bk\!\cdot\!\bOmega \!-\! \bk'\!\cdot\!\bOmega'}$.

For a fixed value of ${ (\bk , \bk' , \bJ) }$,
we must integrate along the resonance line in ${\bJ'}$ given by ${ f_{\bk \bk'} (\bJ , \bJ') \!=\! f (\bJ') \!=\! 0 }$.
To do so, we use the same resonant coordinates, ${ (u,v) }$, as in \extapp~{A1.3} of~\citetalias{Petersen+2024} (see also \extapp~{B} of~\citealt{FouvryPrunet2022}).
We refer to these papers for definitions and
notations.
Within these coordinates, the resonance condition becomes
\begin{equation}
\label{eq:resonance_condition_u_discs}
f (u',v') = \Omega_0 \,\Delta_{\bk'} ( u_{\res} - u' ),
\end{equation}
with the resonant value
\begin{equation}
\label{eq:discs_u_res}
u_{\res} = \varpi_{\bk'}(\bk\!\cdot\!\bOmega/ \Omega_0) ,
\end{equation}
where
\begin{equation}
    \label{eq:varpi_discs}
    \varpi_{\bk'}(\omega) = \frac{\omega - \Sigma_{\bk'}}{\Delta_{\bk'}}.
\end{equation}
In \eqs~(\ref{eq:resonance_condition_u_discs}--\ref{eq:varpi_discs}), we introduced the frequency scale, ${\Omega_0\!=\!V_0/R_0}$, along with ${\Sigma_{\bk'} \!=\! \half(\omega_{\bk'}^{\min} \!+\! \omega_{\bk'}^{\max})}$ and ${\Delta_{\bk'} \!=\! \half (\omega_{\bk'}^{\max} \!-\! \omega_{\bk'}^{\min})}$, where ${\omega_{\bk'}^{\min}}$ (resp.\ ${\omega_{\bk'}^{\max}}$) is the minimal (resp.\ maximal) value reached by the (dimensionless) resonance frequency
${\omega_{\bk'} \!=\! \bk'\!\cdot\!\bOmega / \Omega_0}$.
These extrema can be determined following \extapp~{B} of~\cite{FouvryPrunet2022}.
Performing the change of variables ${\bJ'\!\to\!(u',v')}$ in \eq~\eqref{eq:BL_single_resonance_generic} and using the property ${\dirac(\alpha x)\!=\!\dirac(x)/|\alpha|}$, the flux ultimately reads
\begin{equation}
\label{eq:BL_generic_uv}
\flux_{\bk\bk'}(\bJ) \!=\! H(u_{\res})\!\!
\int_{-1}^{1} \!
\frac{\rd v'}{\Omega_0 \Delta_{\bk'}}
\left| \frac{\p \bJ'}{\p (u',v')} \right|
G_{\bk\bk'}[\bJ,\bJ'(u_{\res},v')] , \notag
\end{equation}
with the rectangular Heaviside function, ${ H\!=\!\mathds{1}_{[-1,1]} }$, imposing ${ |u_{\res}| \!\leq\! 1 }$.
In practice, we performed these integrals using the midpoint rule with 100 sampling points.
Following \extapps~{A4.1} and~{F} of \citetalias{Petersen+2024}, the resonance coordinate $v'$ is chosen to spread more evenly along the resonance line using ${{\tiny\texttt{vmapn}} \!=\! 2}$ (see \exteq~{A18} therein).
The dressed coupling coefficients in the integrand, $G_{\bk\bk'}$ (\eq~\ref{eq:BL_G_Flux}), are computed via \eq~\eqref{eq:dressed_coupling_coefficients_effective} using 100 basis elements.
The response matrix is computed with \linearresponse\@, as detailed in \app~\ref{app:lr_params}.

To compute the relaxation rate in \fig~\ref{fig:discs_BL_NBODY},
we finally need to sum over the pairs of resonance numbers, ${ (\bk,\bk') }$.
For discs, symmetry imposes ${k_{\phi} \!=\! k^{\prime}_{\phi}\!=\!\ell}$, with $\ell$ the considered harmonic number.
In practice, following \citetalias{Fouvry+2015}, we limited ourselves to the
inner Lindblad resonance ${ \bk \!=\! (-1 , 2) }$,
the outer Lindblad resonance ${ \bk \!=\! (1 , 2) }$,
and the corotation resonance ${ \bk \!=\! (0,2)}$,
accounting for the harmonics ${ \ell \!=\! \pm 2 }$.
In practice, we also checked that increasing the number of radial resonances
beyond ${ |k_{r}| \!=\! 1 }$ had a negligible impact on the kinetic prediction.
Finally, the relaxation rate, ${\p F/\p t}$,
is computed from the flux using finite differences with the step distance, ${\delta J_r\!=\!\delta L\!=\!10^{-3}}$.

\section{Linear response implementation}
\label{app:linear-response}

Let us list the main differences between \citetalias{Fouvry+2015} and our work
regarding the implementation of linear response:
(i) \citetalias{Fouvry+2015} used the bi-orthogonal basis from~\cite{Kalnajs1976},
while we used the one from~\cite{CluttonBrock1972}.
This latter basis does not suffer from numerically unstable recurrence relations.
(ii) \citetalias{Fouvry+2015} used only 9 basis functions,
but did not perform any convergence check w.r.t.\ these parameters.\footnote{For example,
we checked that using the same control parameters as~\citetalias{Fouvry+2015} within \linearresponse\@~\citep{Petersen+2024}
leads to the incorrect prediction that the half-mass Mestel disc is unstable.}
In the present work, we used 100 basis elements to obtain \fig~\ref{fig:discs_BL_NBODY}.
We also aimed to ensure the numerical convergence
of the susceptibility coefficients.
(iii) \citetalias{Fouvry+2015} computed the resonant integral from \eq~\eqref{eq:polarisation_matrix}
in pericentre and apocentre compared to our tailored resonant coordinates (see \extapp~{A1.3} of~\citetalias{Petersen+2024} and \extapp~{B} of~\citealt{FouvryPrunet2022}).
But, more worryingly, \citetalias{Fouvry+2015} was not able to evaluate ${ U^{\rd}_{\bk\bk'} (\omega) }$
for purely real frequencies, and resorted to computing it for slightly imaginary frequencies.
Our use of the analytical continuation proposed by~\cite{FouvryPrunet2022}
allowed us to explicitly circumvent this issue.

The combination of all these elements lead us to believe
that the dressed coupling coefficients computed by~\citetalias{Fouvry+2015} were not converged.
Since an incorrect linear response calculation
can lead to an incorrect \BL\ prediction (\fig~\ref{fig:ridges_modes}),
the \BL\ prediction presented in~\citetalias{Fouvry+2015}
should, surely, be considered with great circonspection.

\section{Landau equation}
\label{app:Landau}

In the absence of collective amplification,
\BL\ becomes the Landau equation~\citep[see, e.g.\@,][]{Chavanis2013stellar}.
This amounts to making the replacement ${ U^{\rd}_{\bk\bk'} (\bJ , \bJ' , \omega) \!\to\! U_{\bk\bk'} (\bJ , \bJ')}$
in \eq~\eqref{eq:BL}.
The bare susceptibility coefficients, ${ U_{\bk\bk'} (\bJ , \bJ') }$,
are the Fourier transform in angles
of the pairwise interaction potential, i.e., they read
\begin{align}
U_{\bk\bk'} (\bJ , \bJ') {} & = \!\! \int \!\! \frac{\rd \mybtheta}{(2 \pi)^{d}} \frac{\rd \mybtheta'}{(2 \pi)^{d}} \, U \big[ (\mybtheta , \bJ) , (\mybtheta' , \bJ') \big] \, \re^{- \ri \cdot (\bk \cdot \mybtheta - \bk' \cdot \mybtheta')}
\nonumber
\\
{} & =  - \sum_{p} \psi^{(p)}_{\bk}(\bJ) \, \psi^{(p)*}_{\bk'}(\bJ') 
\label{eq:bare_coupling_coefficients}
\end{align}

In the razor-thin axisymmetric case,
we can expand the interaction potential
in the polar coordinates ${ (r , \phi) }$ to give
\begin{equation}
U ([r , \phi] , [r' , \phi']) = \sum_{\ell} U^{\ell} (r , r') \, \re^{\ri \ell (\phi - \phi')}\,,
\label{eq:intro_Ul}
\end{equation}
with
\begin{equation}
U^{\ell} (r , r') = \frac{1}{\pi} \!\! \int_{0}^{\pi} \!\! \rd \gamma \, U (r , r' , \gamma) \, \cos (\ell \gamma) ,
\label{eq:calc_Ul}
\end{equation}
This last function can be written as
\begin{align}
\nonumber
U^{\ell}_{\bk\bk'} (\bJ,\bJ') = &{}
\frac{\delta^{\ell}_{k_{\phi}}\delta^{\ell}_{k_{\phi}^{\prime}}}{\pi^2} 
\int_0^{\pi} \!\! \rd\theta_r
\!\! \int_0^{\pi} \!\! \rd\theta_r^{\prime} \, 
U^{\ell} (r,r') \\
\nonumber
&\times
\cos\!\left[k_r \theta_r \!+\! k_{\phi}(\theta_{\phi}\!-\!\phi)\right] \\
&\times
\cos\!\left[k_r^{\prime} \theta_r^{\prime} \!+\! k_{\phi}^{\prime}(\theta_{\phi}^{\prime}\!-\!\phi')\right],
\label{eq:bare_coefficients_disc}
\end{align}
For the gravitational case, we have
\begin{equation}
\label{eq:def_U_gamma}
U(r,r',\Delta\phi)=\frac{-G}{\sqrt{r^2+r^{\prime 2}-2 r r' \cos\Delta\phi}}\,,
\end{equation}
from which we find 
\begin{equation}
U^{\ell}(r,r') \!=\! \frac{-G}{\overline{r}}
\frac{_3\tilde{F}_2\!\left[(\half,\half,1),(1\!-\!\ell,1\!+\!\ell),
2a/(1\!+\!a)\right]}{\sqrt{1+a}},
\label{eq:def_F3}
\end{equation}
where ${\overline{r}\!=\!\sqrt{r^2 \!+\! r^{\prime 2}}}$, ${a\!=\!2r r'/\overline{r}^2}$
and ${_3\tilde{F}_2}$ is the regularised (generalised) hypergeometric function.
Equation~\eqref{eq:def_F3} diverges for ${r\!=\!r'}$,
but this divergence is to be integrated over through \eq~\eqref{eq:bare_coefficients_disc}.
In practice, to avoid numerical instabilities,
we regularise the interaction potential by using the Plummer softened potential, $U_{\veps}$~\citep{Aarseth1963,Dehnen2001}
\begin{equation}
\label{eq:Plummer_softening_pairwise}
U_{\veps} (\br , \br') = - \frac{G}{\sqrt{\veps^2 \!+\! \|\br \!-\! \br'\|^2}},
\end{equation}
with ${ \veps / R_{0} \!=\! 10^{-5} }$.
Fortunately, even with softening, ${U^{\ell}_{\veps}(r,r')}$ still follows \eq~\eqref{eq:def_F3}
with the change ${\overline{r}\!=\!\sqrt{r^2 \!+\! r^{\prime 2} \!+\! \veps^2}}$.

In practice, the double integral in \eq~\eqref{eq:bare_coefficients_disc} can be computed
using the midpoint rule. This calculation is useful to benchmark the biorthogonal basis calculations,
by comparing the two expressions from \eq~\eqref{eq:bare_coupling_coefficients}.
Reassuringly, we find
that both approaches are in satisfactory agreement with one another.
In practice, we used \eq~\eqref{eq:bare_coupling_coefficients} with $100$ basis elements to compute the Landau prediction presented in \fig~\ref{fig:discs_longterm_predictions_Landau}.
All the other parameters of the computation
are kept the same as for the \BL\ prediction (\app~\ref{app:BL}).

\section{Softening}
\label{app:discs_stability_softening}

In \Nbody\@ simulations, the Newtonian interaction potential, ${U(r)\!=\!-G/r}$,
diverges at small separations.
As a result, one typically softens the interaction potential via~\citep{Dehnen2001}
\begin{equation}
\label{eq:softening_kernel}
U_{\veps}(r) = \frac{-G}{\veps} f\!\left(\frac{r}{\veps}\right),
\end{equation}
with $f$ a dimensionless kernel and $\veps$ the softening length.

The impact of softening on the stability of self-gravitating systems
has been well studied~\cite[see, e.g.,][]{Miller1971,SaloLaurikainen2000,SellwoodEvans2001,Polyachenko2013}.
Typically, too much softening leads to a strong bias in the gravitational force,
while too little softening leads to numerical instabilities~\citep{Merritt1996}.
This is a classical bias-variance trade-off.

In that context, \citet[][]{DeRijcke+2019b} (hereafter~\citetalias{DeRijcke+2019b})
investigated specifically the impact of the softening kernel
on the instabilities of self-gravitating discs.
In particular, \citetalias{DeRijcke+2019b} considered two softening kernels,
namely the Plummer kernel\footnote{This kernel mimics the effect of finite thickness~\citep{Sellwood2014}.}
\begin{equation}
\label{eq:Plummer_softening}
f_{\mathrm{P}}(u) = \frac{1}{\sqrt{1+u^2}},
\end{equation}
and the (modified) Kuzmin kernel
\begin{equation}
\label{eq:Kuzmin_softening}
f_{\mathrm{K}}(u) = \frac{3+\tfrac{5}{2}u^2+u^4}{(1+u^2)^{5/2}}.
\end{equation}

Interestingly, \citetalias{DeRijcke+2019b} showed that the Plummer softening
impacts the disc's unstable modes at order ${ \mO (\veps) }$,
while the Kuzmin kernel only impacts at order ${ \mO (\veps^{2}) }$.
This makes it a preferred kernel for the razor-thin geometry.
\citetalias{DeRijcke+2019b} compared their theoretical predictions of linear instabilities
with the Plummer softening
against the \Nbody\@ simulations
of~\cite{EarnSellwood1995} (isochrone disc) and~\cite{SellwoodEvans2001} (Mestel disc).
A good agreement was found on the modes' frequencies.

\begin{figure}
\centering
\includegraphics[width=0.48\textwidth]{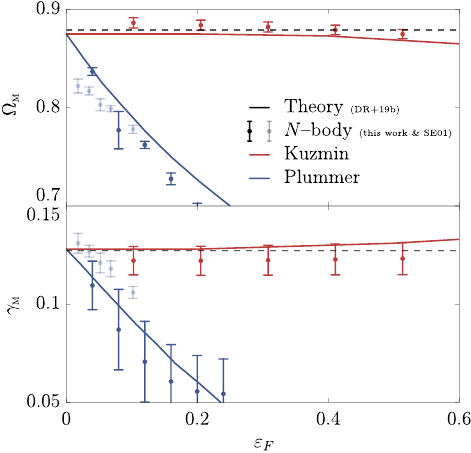}
\caption{Pattern speed (top) and growth rate (bottom) of the dominant mode in Zang ${\nu\!=\!4}$ disc~\citep{ZangThesis},
as a function of the rescaled softening length, $\veps_{F}$ (\eq~\ref{eq:softening_F}), for the Plummer softening kernel (blue -- \eq~\ref{eq:Plummer_softening}) and the Kuzmin kernel (red -- \eq~\ref{eq:Kuzmin_softening}).
The theoretical predictions (plain lines) are reproduced from \citetalias{DeRijcke+2019b} (\extfig~{3} therein), with the dashed line corresponding to the Newtonian (non-softened) prediction from~\citet{EvansRead1998} (\exttab~{D1b} therein).
The \Nbody\ measurements are from~\cite{SellwoodEvans2001} (faint dots SE01 -- Plummer softening only -- \extfig~{1} panel b therein),
and our own measurements (plain dots -- both kernels).
Here, we recover that the Plummer kernel has a strong gravity bias, i.e., a predicted change in the mode's property ${\Delta\omegaM^{\mathrm{Plummer}} \!\propto\! \veps}$,
while the Kuzmin kernel satisfies ${\Delta\omegaM^{\mathrm{Kuzmin}}\!\propto\!\veps^2}$.
}
\label{fig:DeRijcke_modes}
\end{figure}

In this Appendix, we conduct a similar investigation for the Zang ${\nu\!=\!4}$ disc (same as \citealt{SellwoodEvans2001}), using both the Plummer and Kuzmin kernels.
The results are presented in \fig~\ref{fig:DeRijcke_modes}.
On the \Nbody\ front, for each kernel and softening length, we performed 10 independent realisations with ${N\!=\!2\!\times\!10^8}$ particles each.
We found that the unstable modes were indeed less affected by the Kuzmin kernel compared to the Plummer kernel.
Our numerical measurements align with the predictions from~\citetalias{DeRijcke+2019b} as well as the earlier results of~\cite{SellwoodEvans2001}.
However, it is important to note that in this figure, the \Nbody\@ errors are likely underestimated: \cite{SellwoodEvans2001} did not account for the scatter among different realisations,
while we did not include any uncertainties in the mode fits.

In \fig~\ref{fig:DeRijcke_modes}, the softening length, $\veps$, is rescaled.
Indeed, since \eq~\eqref{eq:softening_kernel} is invariant under the transformation \citepalias[\extsect~{2.3} of][]{DeRijcke+2019b} 
\begin{equation}
\label{eq:softening_rescaling}
\veps \to a \, \veps \quad \text{and} \quad f (u) \to a f (a \, u) ,
\end{equation}
one ought to be careful when comparing softening lengths.
Here, we follow~\citetalias{DeRijcke+2019b},
and compare the kernels using
\begin{equation}
\label{eq:softening_F}
\veps_F = - \veps / f'_{\max} ,
\end{equation}
with ${f'_{\max}}$ the maximal value of the derivative of the softening (dimensionless) kernel.
Setting ${ \veps_F / \veps \!=\! 1 }$ for the Plummer softening, one has ${ \veps_F / \veps \!\approx\! 2.568 }$ for the Kuzmin softening \citepalias[\exttab~{1} in][]{DeRijcke+2019b}.

\end{document}